\documentclass[fleqn,usenatbib]{mnras}

\usepackage{newtxtext,newtxmath}
\usepackage[T1]{fontenc}
\usepackage{color}
\usepackage[dvipsnames]{xcolor}

\DeclareRobustCommand{\VAN}[3]{#2}
\let\VANthebibliography\thebibliography
\def\thebibliography{\DeclareRobustCommand{\VAN}[3]{##3}\VANthebibliography}

\usepackage{graphicx}	% Including figure files
\usepackage{amsmath}	% Advanced maths commands
\usepackage{soul}
\sethlcolor{red}

\newcommand{\red}[1]{}
\newcommand{\ty}[1]{}
\newcommand{\high}[1]{}

\newcommand{\msun}{\,\textrm{M}_{\odot}}

\newcommand{\mbar}{M_{\text{bar}}}

\newcommand{\vmax}{V_{\text{max}}}
\newcommand{\vflat}{V_{\text{flat}}}
\newcommand{\vexp}{V_{\text{2.2}}}
\newcommand{\rexp}{R_{\text{2.2}}}
\newcommand{\wpt}{\text{W}_{\text{p20}}}
\newcommand{\kl}{D_\text{KL}}
\newcommand{\mtot}{M_{\text{tot}}}
\newcommand{\rout}{R_{\text{out}}}
\newcommand{\vout}{V_{\text{out}}}
\newcommand{\rmax}{R_{\text{max}}}
\newcommand{\reff}{R_{\text{eff}}}
\newcommand{\vbar}{V_{\text{bar}}}
\newcommand{\vobs}{V_{\text{obs}}}

\newcommand{\mvir}{M_{\text{halo}}}

\newcommand{\HI}{\ensuremath{\mathrm{H}\scriptstyle\mathrm{I}}}

\newcommand{\dkl}{D_{\text{KL}}}

\title[Where is the information on halo properties?]{The information on halo properties contained in spectroscopic observations of late-type galaxies}
\author[T. Yasin et al.]
{Tariq Yasin,$^{1}$\thanks{\href{mailto:tariq.yasin@physics.ox.ac.uk}{tariq.yasin@physics.ox.ac.uk}} 
Harry Desmond,$^{2}$
Julien Devriendt,$^{1}$
and Adrianne Slyz$^{1}$
\\ 
$^1$Astrophysics, University of Oxford, Denys Wilkinson Building, Keble Road, Oxford, OX1 3RH, UK\\
$^2$Institute of Cosmology \& Gravitation, University of Portsmouth, Dennis Sciama Building, Portsmouth, PO1 3FX, UK\\
}

\date{Accepted XXX. Received YYY; in original form ZZZ}

\pubyear{2023}

\begin{document}
\label{firstpage}
\pagerange{\pageref{firstpage}--\pageref{lastpage}}
\maketitle

\begin{abstract}
Rotation curves are the key observational manifestation of the dark matter distribution around late-type galaxies. In a halo model context, the precision of constraints on halo parameters is a complex function of properties of the measurements as well as properties of the galaxy itself. Forthcoming surveys will resolve rotation curves to varying degrees of precision, or measure their integrated effect in the \HI{} linewidth. To ascertain the relative significance of the relevant quantities for constraining halo properties, we study the information on halo mass and concentration as quantified by the Kullback--Leibler divergence of the kinematics-informed posterior from the uninformative prior. We calculate this divergence as a function of the different types of spectroscopic observation, properties of the measurement, galaxy properties and auxiliary observational data on the baryonic components. Using the SPARC sample, we find that fits to the full rotation curve exhibit a large variation in information gain between galaxies, ranging from \textasciitilde{1} to  \textasciitilde{11} bits. 
The variation is predominantly caused by the vast differences in the number of data points and the size of velocity uncertainties between the SPARC galaxies. We also study the relative importance of the minimum \HI{} surface density probed and the size of velocity uncertainties on the constraining power on the inner halo density slope, finding the latter to be significantly more important. 
We spell out the implications of these results for the optimisation of galaxy surveys aiming to constrain galaxies' dark matter distributions, highlighting the need for precise velocity measurements.
\end{abstract}

\begin{keywords}
dark matter -- galaxies: kinematics and dynamics -- galaxies: statistics
\end{keywords}

%%%%%%%%%%%%%%%%%%%%%%%%%%%%%%%%%%%%%%%%%%%%%%%%%%

%%%%%%%%%%%%%%%%% BODY OF PAPER %%%%%%%%%%%%%%%%%%

\section{Introduction}\label{sec:introduction}

In the standard model of cosmology, dark matter clusters under the action of gravity to form virialised, approximately spherical structures known as haloes. Galaxy formation occurs when baryons fall into their deep potential wells, cooling and condensing to form stars. Constraining the relationship between galaxies and their dark matter haloes, the ``galaxy--halo connection'', is an important step to constructing a complete theory of galaxy formation \citep[see][and references therein]{wechslerConnectionGalaxiesTheir2018}. 

Perhaps the most direct method to measure dark matter halo density profiles is through spectroscopic observations of galaxies' kinematics. The mass distribution is then inferred by comparing the observed motions of the gas and stars to the motions expected from the observed luminous matter through Newtonian gravity. Late-type galaxies are ideal systems to study, as their stars and gas follow approximately circular orbits that directly trace the underlying potential. Spatially resolved observations are able to measure rotation curves (RCs), the rotational velocity of the stars and gas as a function of radius (e.g.~\citealt{walterTHINGSHINearby2008,ponomarevaDetailedHIKinematics2016,lelliSPARCMASSMODELS2016}). %, from which the dark matter distribution can be simply derived \citep{katzTestingFeedbackmodifiedDark2017, liComprehensiveCatalogDark2020}. \ty{Martin et. al in prep :)}

It has been argued
% However, it is argued
that observed correlations between the dark matter and baryons implied by the RCs \citep[e.g.][]{mcgaughRadialAccelerationRelation2016,mcgaughDynamicalRegularitiesGalaxies2019} cannot be explained in the cold dark matter paradigm. This has lead to an abundance of proposed extensions to the standard model, either in the form of new dark matter physics \citep[e.g.][]{khelashviliDarkMatterProfiles2022,adhikariAstrophysicalTestsDark2022a} or modifications to the theory of gravity \citep[e.g.][]{milgromModificationNewtonianDynamics1983,burrageRadialAccelerationRelation2017,naikConstraintsChameleonGravity2019}. Improving the precision of constraints on halo properties is therefore also vital to assess the viability of dark matter models. One approach is to compare the halo properties inferred from kinematics to cosmological expectations \citep[e.g][]{katzTestingFeedbackmodifiedDark2017,liComprehensiveCatalogDark2020,postiDynamicalEvidenceMorphologydependent2021,mancerapinaImpactGasDisc2022,yasinInferringDarkMatter2022}.

The density profile of dark matter is usually constrained by fitting parameterised functions~\citep[e.g.][]{navarroUniversalDensityProfile1997,burkertStructureDarkMatter1995,dicintioDependenceDarkMatter2014,readDarkMatterCores2016} to the measured RC. The parameters, which we refer to as the \emph{properties} of the halo, are typically the virial mass, the concentration (a measure of the autocorrelation of dark matter within the halo) and sometimes additional parameters describing the shape of the profile, such as the steepness of the inner slope. The tightness of the constraints on the free parameters is a complex function of the properties of the measurements (e.g. the number of measured RC data points, the uncertainties on the measured velocities, the maximum radius probed), properties of the galaxy (e.g. stellar surface density, gas fraction) and the precision of auxiliary data on the galaxy (inclination, distance, baryon content). 

Rotation curves can be measured using a variety of kinematic tracers. Radio telescopes, in either single dish or interferometer configurations, probe atomic hydrogen (\HI{}), which has the advantage of extending far beyond the optical disc for late-type galaxies, into the region where dark matter dominates.
% and there is no degeneracy with the baryonic contribution to the circular velocity (the \emph{disk-halo} degeneracy).
The SPARC \citep{lelliSPARCMASSMODELS2016} database, the largest of its type, contains \HI{} rotation curves for 175 late-type galaxies. Collated from archival observations, the measurement properties vary greatly between different RCs. For example, the best sampled RC has 120 data points, and the worst sampled has 5. In this paper we use SPARC to study the precision of constraints on halo parameters as a function of measurement and galaxy properties. This will reveal the ways in which future surveys ought to be designed to maximise their constraining power on the dark matter distributions around galaxies.

Other tracers do not probe as far as \HI{}, but have their own advantages. Optical IFU surveys such as DiskMass \citep{martinssonDiskMassSurveyVI2013}, MaNGA \citep{bundyOverviewSDSSIVMaNGA2014} and SAMI \citep{croomSAMIGalaxySurvey2021} use optical emission from ionised gas or stars, which extends only as far as the galaxy's stellar component, although the resolution is higher than \HI{} surveys. Sub-mm telescopes (again either in single dish or interferometric configurations) can be used to probe the molecular CO gas disk, which usually extends to 50-70\% of the radius of the stellar disc \citep[e.g.][]{langPHANGSCOKinematics2020}.
%(see also the PROBES RC compendium, \citealt{stonePROBESICompendiumDeep2022})

Future and current surveys \HI{} are seeking to study larger, statistically representative samples of galaxies \citep[e.g.][]{maddoxMIGHTEEHIHIEmission2021,oosterlooLatestApertif2010a} and galaxies at larger distances. For example, MIGHTEE-HI is an ongoing blind survey that will measure RCs out to $z=0.5$. Due to the further distances, the RCs are sampled more coarsely than in nearby catalogues such as SPARC \citep{ponomarevaMIGHTEEHBaryonicTully2021}. RCs can be measured to lower radius but higher redshift using optical \citep{stottKMOSRedshiftOne2016,diteodoroFlatRotationCurves2016} and sub-mm \citep{jonesALPINEALMAIiSurvey2021,lelliMassiveStellarBulge2021,lelliColdGasDisks2023} instruments. Strongly lensed galaxies can also be used to increase the distance probed \citep[e.g.][]{rizzoDynamicalPropertiesDusty2021}.

Finally all of the methods described above can be used to obtain spatially integrated spectra, which requires significantly less integration time and so can be obtained for orders of magnitude more galaxies. Current surveys such as ALFALFA \citep{haynesAreciboLegacyFast2018}, and future surveys such as WALLABY \citep{koribalskiWALLABYSKAPathfinder2020,degWALLABYPilotSurvey2022a}, CRAFTS \citep[][undertaken on FAST]{zhangPredictionsFASTTelescope2020} and the SKA \citep{yahyaCosmologicalPerformanceSKA2015} will measure the spatially integrated \HI{} emission for cosmological volumes of galaxies.

These different observational techniques and tracers will measure rotation velocities with varying precision and will probe different parts of the RC. The constraining power of these different types of kinematic measurement on halo properties has never been studied in detail. To begin investigation of this question, we study here
%Therefore we also study
the information content contained in different parts of the RCs of the SPARC galaxies. This is achieved by constraining halo properties using the RC summary statistics recorded in SPARC, each of which characterise a particular part of the RC. These statistics are $\vflat$, the speed of the flat part of the rotation curve, $\vmax$, the peak rotational speed and $\vexp$, the circular velocity at 2.2 times the disc exponential scale length and $\wpt$ the width of the spatially integrated \HI{} line profile at 20\% of the peak flux.

RC summary statistics have previously been utilised extensively in the study of the Tully--Fisher relation (TFR), the tight empirical relationship between the mass or luminosity of a galaxy and some measure of its rotational velocity. The TFR was first defined using \HI{} linewidth \citep{tullyReprint1977A541977}, and later studies have used $\vmax$ \citep{conseliceRelationshipStellarHalo2004}, $\vexp$ \citep{courteauOpticalRotationCurves1997} or $\vflat$ \citep{mcgaughBaryonicTullyFisherRelation2005,ponomarevaLightBaryonicMass2018}. Recently \cite{lelliBaryonicTullyFisherRelation2019} (henceforth L19) compared the baryonic TFR \citep{mcgaughBaryonicTullyFisherRelation2000,mcgaughBaryonicTullyFisherRelation2005,lelliSMALLSCATTERBARYONIC2016,iorioLITTLETHINGS3D2016a} produced by different summary statistics of the RC, as well as the \HI{} line width, for the SPARC galaxies. The study of the dark matter halo constraints offered by these summary statistics is also interesting in this context.

A velocity summary statistic is sensitive only to the enclosed dynamical mass within the radius it probes (which can be estimated as $M\sim V^2R/G$, \citealt{castertanoGALAXYMASSESDETERMINED1980}). Hence, although summary statistics do provide information on halo mass and concentration (once a profile is assumed), the constraints are relatively weak due to the degeneracy between the two parameters: the same enclosed mass can be generated by either a high mass, low concentration or a low mass, high concentration halo. The degeneracy can be broken to some extent by assuming a halo mass--concentration relation from simulations \citep[e.g.][]{posti2019}, but the extent to which these relationships are obeyed by real halos is uncertain \citep{duttonColdDarkMatter2014,katzTestingFeedbackmodifiedDark2017,liComprehensiveCatalogDark2020,mancerapinaImpactGasDisc2022}. For example, due to assembly bias \citep{dalalHaloAssemblyBias2008}, we do not expect a population of relatively isolated, late-type galaxies such as those found in SPARC to follow an identical mass--concentration relationship to that of all the halos in a simulation. 

Without an informative prior linking mass and concentration, the kinematic data alone can sometimes not exclude unphysical scenarios such as low mass galaxies having cluster mass halos. Studying the degenerate posterior in mass--concentration can still yield insight however. The constraints from summary statistics can be compared and/or combined with other pieces of information, such as abundance matching \citep{yasinInferringDarkMatter2022}, optical kinematics \citep[proposed by][]{taranuSelfconsistentBulgeDisk2017a} or weak lensing data \citep[][for SDSS velocity dispersions]{shajibDarkMatterHaloes2021}. Therefore in this study we choose to analyse the constraining power in the mass--concentration plane offered by the different types of measurements without applying a mass--concentration prior, although we discuss the effects this would have in Sec.~\ref{sec:summ_stat}. % individual  summary statistics/integrated quantities.

We quantify the precision of constraints on halo properties using the Kullback--Leibler divergence \citep[$\dkl$,][]{kullbackInformationSufficiency1951} of the posterior from the prior, a measure of information gain based on information theory. We study the information gain of the 2D total mass--concentration posterior (where total mass is equal to the halo mass plus the galaxy mass). We use total mass rather than halo mass, because the lower bound of the prior on total mass can be naturally set to the galaxy mass, whereas the halo mass, when sampled logarithmically (as is computationally necessary), has no natural lower bound. Our method could equally be applied to additional parameters describing shape. By quantifying the information gain on these halo properties when using either the full RCs, \HI{} line widths or summary statistics to constrain the kinematic model, we aim to answer the following questions:

  \begin{itemize}
  \item How does the information content depend on properties of the measurement such as velocity uncertainties, the minimum \HI{} surface density probed and auxiliary data on galaxy parameters.
  \item How does the information content depend on galaxy properties?
  \item How much information is contained in different summary statistics compared to the full RC?
  \item How much information is there in integrated \HI{} measurements relative to more expensive, spatially resolved measurements?% \purple{that require interferometry}?
  \end{itemize}

To answer the first two questions we will construct a predictive model for $\dkl$ given the galaxy and measurement properties as input. The paper is structured as follows. Section~\ref{data} describes the the SPARC data. Section~\ref{methods} describes the Bayesian models for inferring halo properties from the observations. In Section~\ref{results} we present the information content as a function of different types of observations, measurement properties and galaxy properties. We discuss the implications of our results in Section~\ref{discussion}, and conclude in Section~\ref{conclusion}. We define the halo mass $\mvir$ using the overdensity condition \(\Delta_{\text{vir}}=178\) of  \citet{bryanStatisticalPropertiesXray1998}. All logarithms are base-10 unless stated otherwise. %We assume \(H_0 = 70 \kms\), and .

\section{Observational data}\label{data}

\begin{table*}
\caption{\label{tab:definitions}The summary statistics of the full rotation curve and \HI{} linewidth definition used in this paper.} 
\begin{tabular}{ |c|p{8cm}|c| }
  \hline
  \textbf{Measurement} & \textbf{Interpretation} & \textbf{{Sample size}} \\
  \hline
   Full RC & The full rotation curve. & 175 \\
  \hline
   $\vflat$ & 
   The velocity of the flat part of the rotation curve. The algorithm to identify the flat part of the rotation curve is listed in Section \ref{sec:vflat} and differs slightly from the definition used in the SPARC database. &  123\\
  \hline
   $\vmax$ & 
   The maximum velocity of the rotation curve. If the rotation curve is continuously rising in the observed range then this is the outermost measured point & 175\\
  \hline
   $\vexp$ & 
   The velocity at twice the exponential stellar disk scale length. This is found by linearly interpolating between datapoints on either side of the required radius. & 167\\
  \hline
   $\text{W}_{\text{p20}}$ & 
   The width of the global \HI{} 21-cm emission line of a galaxy measured at 20\% of the peak flux. & 148 \\ 
    \hline

\end{tabular}
\end{table*}

SPARC\footnote{\url{http://astroweb.cwru.edu/SPARC/}} \citep{lelliSPARCMASSMODELS2016} is a database of rotation curves for 175 late-type galaxies, collated from archival resolved \HI{} observations. In addition, 56 galaxies have hybrid rotation curves with high resolution $H\alpha$ data in the inner parts. Each galaxy has Spitzer photometry \citep{schombertStellarPopulationsStar2014} at $3.6 \mu\text{m}$, a band in which the mass-to-light ratio is relatively constant \citep{schombertMasstolightRatiosStar2018}, which reduces the disk-halo degeneracy in kinematic analyses. The distributions of the stars and gas (\emph{mass models}) are provided in the form of the contribution of each component to the circular velocity as a function of radius. We also utilise the \HI{} surface density ($\Sigma_{\text{HI}}$) as a function of radius (F. Lelli, private communication) in our analysis. The SPARC galaxies span a wide range in luminosity  (\(10^7\) to \(10^{12} L_{\odot}\)), surface brightness (\textasciitilde5 to \textasciitilde{}\(5000 \: L_{\odot} \text{pc}^{-2}\)), \HI{} mass (\textasciitilde{}\(10^7\) to \(~10^{10.6} M_{\odot}\)) and morphological type (S0 to Im/BCD).

L19 calculate the RC summary statistics $\vflat$, $\vmax$, $\vexp$ and their associated errors for a subset of SPARC sample with cuts on properties such as $i$ and number of data points. We use their definitions to calculate $\vmax$ and $\vexp$ for the whole sample, but we calculate $\vflat$ using our own definition, which we describe in Section~\ref{sec:vflat}. The definitions of the velocity measurements are summarised in Table~\ref{tab:definitions}. L19 also compile $\HI{}$ linewidths from archival data for various different definitions. We choose to use $\wpt$, the width at 20\% of the peak flux, as it is available for the most galaxies. Unlike L19, we do not include the contribution from inclination to the observational uncertainties on the summary statistics and $\wpt$, as inclination is a free parameter in our inference.

\section{Methods}\label{methods}

\begin{table*}
\caption{\label{tab:parameters}The free parameters in our kinematic model, their physical definitions and their Bayesian priors. We sample all parameters in logarithmic space except inclination and distance.}
\begin{tabular}{ |c|c|c|c|c| }
  \hline
   & \textbf{Parameter} & \textbf{Units} & \textbf{Definition} & \textbf{Prior} \\

  \hline
   & $M_{\text{tot}}$ & $M_{\odot}$ & Total mass $M_{\text{tot}} = M_{\text{halo}} + M_{\text{bar}}$ & Flat in range $\log(\mbar/\msun) < \log(M_{\text{tot}} / \msun) < 15.5$ \\
   & $c_{0.1}$ &  -  & Halo concentration, as defined in Eq.~(\ref{eq:conc}) & Flat in range $0.5 < \log c_{0.1} < 2$\\
   & $\Upsilon_{\text{disc}}$ & $\msun / \text{L}_{\odot}$ & Disc  mass-to-light ratio & Lognormal ($\mu=\log(0.5)$,$\sigma=0.1$) \\
   & $\Upsilon_{\text{bulge}}$ & $\msun / \text{L}_{\odot}$ & Bulge mass-to-light ratio & Lognormal ($\mu=\log(0.7)$,$\sigma=0.1$) \\
   & $D$ & Mpc & Physical distance to galaxy & Gaussian prior from SPARC value and its uncertainty\\
   & $i$ & deg & Inclination ($0^\circ $ face on; $90^\circ$ edge on)  & Gaussian prior from SPARC value and its uncertainty  \\
  \hline

\end{tabular}
\end{table*}

\subsection{Rotation curve model}

Dark matter halo properties are inferred by fitting a parameterised halo profile to the observational data. Different halo profiles has been studied extensively in literature but a clear picture has yet to emerge of the relationship between the properties of a galaxy and the shape of its halo \citep{katzTestingFeedbackmodifiedDark2017,liComprehensiveCatalogDark2020}. The haloes in dark matter-only simulations were found to have a universal profile dubbed Navarro-Frenk-White (NFW) \citep{navarroUniversalDensityProfile1997}, but profiles motivated by hydrodynamical simulations, that interpolate between a cusp and a core based on galaxy/halo properties \citep{dicintioDependenceDarkMatter2014,readDarkMatterCores2016}, have been found to fit the SPARC data better than the NFW profile \citep{katzTestingFeedbackmodifiedDark2017,liComprehensiveCatalogDark2020}. However cores have been observed in high mass systems \citep{collettCoreCuspsCentral2017}, including for the SPARC galaxies \citep{liConstantCharacteristicVolume2019,liIncorporatingBaryondrivenContraction2022}, which is against the mass-dependant behaviour of the supernova-induced core flattening predicted by the aforementioned hydrodynamical simulations. On the other hand many studies have found NFW haloes to be good fits to clusters and weak lensing data for high mass galaxies (e.g.~\citealt{mandelbaumStrongBimodalityHost2016}), and some have argued that the inference of cores from 21-cm rotation curve observations may be due to systematics \citep{roperDiversityRotationCurves2022}. In light of this uncertainty we choose to study the NFW and Burkert \citep{burkertStructureDarkMatter1995} profiles as representative examples of a cusped and cored profiles respectively. Due to the preference of most SPARC galaxies for cores using our modelling procedure (in agreement with \citealt{liComprehensiveCatalogDark2020}, although see \citealt{posti2019} for a different analysis), we present our results primarily for the Burkert profile.

The NFW density profile is
\begin{equation}
\rho_{\mathrm{NFW}}(r)=\frac{\rho_{s}}{\left(\frac{r}{r_{s}}\right)\left[1+\left(\frac{r}{r_{s}}\right)\right]^{2}},
\label{eq:nfw}\end{equation}
where $r_s$ is a scale radius and $\rho_{s}$ a characteristic density. The enclosed mass at radius \(r\) is
\begin{equation}{
M_{\mathrm{NFW}}(r) =4 \pi \rho_s r_{s}^{3}\left[\ln (1+x)-\frac{x}{1+x}\right]
,}\end{equation}
where \(x\equiv r/r_s\).
%\textbf{Burkert:}  %Unlike the PISO, the enclosed mass of the Burkert profile does not diverge at large radii and is instead the same as the NFW profile. 
The Burkert density profile is
\begin{equation}\protect\hypertarget{eq:burkert}{}{
\rho_{\text {Burkert }}(r)=\frac{\rho_s}{\left(1+\frac{r}{r_{s}}\right)\left[1+\left(\frac{r}{r_{s}}\right)^{2}\right]},
}\label{eq:burkert}\end{equation}
and the enclosed mass is given by
\begin{equation}\protect{}{
M_{\text {Burkert }}(r)=2 \pi \rho_s r_{c}^{3}\left[\frac{1}{2} \ln \left(1+x^{2}\right)+\ln (1+x)-\arctan (x)\right].
}\end{equation}

We also analyse how well observations can constrain a shape parameter for the inner halo by studying the generalised-NFW profile (gNFW, e.g. \citealt{umetsuPreciseClusterMass2011}), with density

\begin{equation}
\rho_{\mathrm{gNFW}}(r)=\frac{\rho_{s}}{\left(\frac{r}{r_{s}}\right)^{\alpha}\left[1+\left(\frac{r}{r_{s}}\right)\right]^{3-\alpha}}.
\label{eq:gnfw}\end{equation}
This reduces to NFW for $\alpha=1$. The mass enclosed is given by

\begin{equation}
M_{\mathrm{gNFW}}(r) =4 \pi \rho_s r_{s}^{3}\left[ B(x/(1+x),3-\alpha,0) \right],
\end{equation}
where $B(z ; a, b) \equiv \int_0^z u^{a-1}(1-u)^{b-1} d u$. 

Typically rotation curve measurements extend to only a small fraction of the virial radius \citep[e.g.][]{katzTightEmpiricalRelation2019}. Therefore constraining halo mass requires a large extrapolation. Whilst gNFW, NFW and Burkert differ in shape towards the centre, at large radii they all decline as \(\rho \propto 1/r^3\), so the comparison between the inferred masses is fair.

The circular speed due to the dark matter at radius \(r\) is \(V_{\text{DM}} = \sqrt{GM_{\text{DM}}(r)/r}\).
It is conveniently expressed in terms of $M_{\text{halo}}$, the virial velocity ($V_{\text{halo}}$) and the virial radius ($R_{\text{halo}}$),
\begin{equation}\protect{}{
\frac{V_{\text{DM}}(r)}{V_{\text{halo}}} = \sqrt{\frac{M_{\text{DM}}(r)}{M_{\text{halo}}} \frac{R_{\text{halo}}}{r}},
}\end{equation}
where $V_{\text{halo}} \equiv \sqrt{G M_{\mathrm{halo}}/R_{\mathrm{halo}}}$.
Concentration is commonly defined based on the radius at which the logarithmic slope of the density profile is -2 ($r_{-2}$, which for NFW is equal to $r_s$)
\begin{equation}\protect{}{
c = \frac{R_{\text{halo}}}{r_{-2}}.
}\end{equation}
Following \citet{yasinInferringDarkMatter2022} we instead use the definition 

\begin{equation}\protect\hypertarget{eq:conc}{}{
c_{0.1} = \frac{R_{\text{halo}}}{r_{0.1}},
}\label{eq:conc}\end{equation}
where \(r_{0.1}\) is the radius enclosing 10\% of $M_{\text{halo}}$. This definition has three advantages: 1) it can be calculated in simulations by simply counting dark matter particles, rather than fitting a profile; 2) $r_{0.1}$ is defined for all halo profiles, as it does not require a particular slope; 3) it is not based on the halo's shape, and hence is more intuitive.

Our Bayesian fitting procedure is similar to those of \citet{katzTestingFeedbackmodifiedDark2017,liComprehensiveCatalogDark2020}. As the late-type galaxies studied are rotationally dominated, we assume the rotational speed is equal to the circular speed. We try a non-fiducial model where we add $10 \text{km/s}$ in quadrature to all velocities as a crude "asymmetric drift" correction, and find it does not affect our results (see \citealt{bureauEnvironmentRamPressure2002,ohHIGHRESOLUTIONMASSMODELS2015,iorioLITTLETHINGS3D2016a} for a full discussion of asymmetric drift). The total circular speed, $V_\text{c}(r)$ is equal to the sum in quadrature of the circular speed due to the dark matter and each baryonic component (dark matter, gas, stellar disc, stellar bulge)

\begin{equation}\protect\hypertarget{eq:ph}{}{
V_{\mathrm{c}}^{2}(r)=V_{\mathrm{DM}}|V_{\mathrm{DM}}|+ \Upsilon_{\text{bulge}}V_{\mathrm{bulge}}|V_{\mathrm{bulge}}|  +  \Upsilon_{\text{disc}}V_{\mathrm{disc}}|V_{\mathrm{disc}}|
 + V_{\mathrm{gas}}|V_{\mathrm{gas}}|,
}\label{eq:ph}
\end{equation}
%\Upsilon_{\mathrm{disc}} 
 %\Upsilon_{\mathrm{bulge}} V_{\mathrm{bulge}}^{2}(r)+V_{\mathrm{gas}}^{2}(r)
where each $V$ is also a function of $r$, and $\Upsilon_{\text{bulge/disc}}$ is the mass-to-light ratio of the bulge or disc. The latter are tabulated in the SPARC database for each galaxy. The baryonic mass models depend on the assumed distance as
\begin{equation}
V_{\text{disc,bulge,gas}}(r) \propto \sqrt{D},
\end{equation} 
and the radius depends on the assumed distance as
\begin{equation}
r \propto D.
\end{equation}
The model prediction for the line-of-sight rotational speed is found by correcting $V_{\text{c}}(r)$ for the inclination $i$ of the galaxy ($i=0^{\circ}$ face-on; $i=90^{\circ}$ edge-on)
\begin{equation}
V_{\text{pred}}(r) = V_{\text{c}}(r) \sin i.
\end{equation}
$D,i,\Upsilon{\text{disk}}$ and $\Upsilon_{\text{bulge}}$ are free parameters in the inference. When fitting to the full RC, $V_{\text{pred}}(r)$ can be compared directly to the observed RC. For the summary statistics, $V_{\text{pred}}(r)$ is evaluated at the same radii as the observed data points, and then the same algorithm that was used to calculate each summary statistic from the observed RC is applied.

\subsection{Definition of $V_{\text{flat}}$}
\label{sec:vflat}

The algorithm to calculate $\vflat$ in the SPARC database (see \citealt{lelliSMALLSCATTERBARYONIC2015}) starts by defining the outermost observed data point as being the flat part of the RC, and then adds additional points to it iteratively. At each step the next innermost data point at radius $r_{i-1}$ is added if its speed is within 5\% of the mean of the data points already included: 

\begin{equation}
\Delta  \equiv \frac{|\overline{V} - V_{i-1} |}{\overline{V}} < 0.05.
\end{equation}
If the difference is greater than 5\% the process terminates, and $\vflat$ is the mean of the points already included. A galaxy is only considered to have a defined $\vflat$ if the flat part constitutes at least three points when the algorithm terminates. The definition depends on the distance between the points, which means finely sampled RCs can still be considered flat even if they are much steeper than less finely sampled RCs. 
%To lessen this bias we normalise $\Delta V$ by the difference in radius between the innermost current point ($R_i$) and the point to be added ($R_{i-1}$), in units of $R_{\text{disc}}$, and choose a new condition so a similar number of galaxies are defined to have a flat part
To lessen this bias we change the condition to a limit on $\Delta$ per stellar disk scale length ($R_{\text{disc}}$),

\begin{equation}
\frac{\Delta }{(R_{i} - R_{i-1})/R_{\text{disc}}} < 0.10.
\end{equation}
The condition is set to 10\% per disc scale length so $\vflat$ is defined for a similar number of galaxies (123) as the original definition (133). Our results are not sensitive to the exact value. For galaxies which meet both the old and new $V_{\text{flat}}$ definition, the difference between the two values is negligible. 

\subsection{\HI{} linewidth model}

The summary statistics $\vmax$,$\vexp$ and $\vflat$ can be calculated from the RC alone. To calculate a model $\wpt$ that can be compared to the observed value, we must calculate a model \HI{} integrated spectrum from the RC and the \HI{} surface density profile. Our method is described and validated in detail in \cite{yasinInferringDarkMatter2022}, but we describe it here in brief. We use the method of \citet{obreschkowSIMULATIONCOSMICEVOLUTION2009}, treating the \HI{} disc as a series of concentric infinitely thin rings each with circular velocity given by $V_{\text{pred}}(r)$. We want to calculate the flux observed at the wavelength $\lambda$ that corresponds to gas with a radial velocity $V_{\lambda}$ relative to the kinematic centre of the galaxy. The normalised flux from a single infinitely thin ring of gas is given by

\begin{equation}
\tilde{\psi}\left(V_{\lambda}, V_{\mathrm{pred}}\right)=\left\{\begin{array}{ll}
\frac{1}{\pi \sqrt{V_{\mathrm{pred}}^{2}-V_{\mathrm{\lambda}}^{2}}} & \text { if }\left|V_{\mathrm{\lambda}}\right|<V_{\mathrm{pred}} \\
0, & \text { otherwise. }
\end{array}\right.
\end{equation}
We assume the gas has a constant velocity dispersion of 10km/s (based on observations of local galaxies: \citealt{leroyStarFormationEfficiency2008a,mogotsiHICOVelocity2016}), which broadens the flux distribution from each ring

\begin{equation}
\psi\left(V_{\mathrm{\lambda}}, V_{\mathrm{pred}}\right)=\frac{\sigma_{\HI{}}^{-1}}{\sqrt{2 \pi}} \int^{+\infty}_{-\infty} d V \exp \left[\frac{\left(V_{\mathrm{\lambda}}-V\right)^{2}}{-2 \sigma_{\HI{}}^{2}}\right] \tilde{\psi}\left(V, V_{\mathrm{pred}}\right).
\end{equation}
The total, integrated flux profile of the galaxy is then obtained by integrating across the whole \HI{} disc:

\begin{equation}\protect\hypertarget{eq:convolution}{}{
\Psi_{\mathrm{HI}}\left(V_{\lambda}\right)=\frac{2 \pi}{M_{\mathrm{HI}}} \int_{0}^{\infty} dr \: r \Sigma_{\mathrm{HI}}(r) \psi\left(V_{\lambda}, V_{\mathrm{pred}}(r)\right).
}\label{eq:convolution}\end{equation}
The resulting flux profile is symmetric, so the model $W_{\text{p20}}$ summary statistic can be trivially calculated by finding a peak of the distribution, and moving outwards to larger speed until until the flux drops to 20\% of the maximum. Although the observed profile may be asymmetric, the important quantity is the width, so this should not bias the results. This is verified for the SPARC sample in \cite{yasinInferringDarkMatter2022}.

\subsection{Inference}

Bayes' theorem is used to calculate the probability of our  parameters \(\theta\) condition on our  data $D$ given the model \(\mathcal{M}\),

\begin{equation}
\mathcal{P}(\theta |D,\mathcal{M})=\frac{\mathcal{L}(D|\theta,\mathcal{M})  \pi(\theta|\mathcal{M})}{p(D|\mathcal{M})},
\end{equation}
where  $\mathcal{L}(D|\theta,\mathcal{M})$ is the likelihood of the data, $\pi(\theta|\mathcal{M})$ is the prior probability density and $p(D|\mathcal{M})$ the marginalised likelihood. For {$\vflat$,$\vmax$,$\vexp$,$\wpt$}, which consist of a single observation, the likelihood of the data is

\begin{equation}\protect{}{
\mathcal{L}(W_{\text{obs}}|\theta,\mathcal{M}) = \frac{\exp\{ 
-(W_{\text{obs}} - W_{\text{pred}})^2 / (2\delta W_{\text{obs}}^2) \}}
{\sqrt{2\pi}\delta W_{\text{obs}}} },
\end{equation} where $W_{\text{obs}}$ is the observed velocity summary statistic and $W_{\text{pred}}$ the model prediction. For the full rotation curve the likelihood is

\begin{equation}\label{eq:likelihood}\protect{}{ 
\mathcal{L}(D|\theta,\mathcal{M}) = {\displaystyle \prod_{i}} \frac{\exp\{ 
-(V_{\text{i,obs}} - V_{\text{pred}}(r_\text{i}))^2 / (2\delta V_{\text{i,obs}}^2) \}}
{\sqrt{2\pi}\delta V_{\text{i,obs}}} }.
\end{equation}

When fitting with summary statistics we find some galaxies have non-zero posterior probability at \(M_{\text{halo}}=0\). Therefore in order to raise the lower limit of the posterior to a finite value we sample \(\log{M_{\text{tot}}} = \log(M_{\text{halo}} + \mbar)\) rather than $M_\text{halo}$ itself, setting the lower bound on its flat prior to be $\log\mbar$. A minimum baryonic content for each galaxy regardless of $\Upsilon$ is ensured by the \HI{} mass, which is relatively well constrained by observations and so is not allowed to vary in our model (apart from through its dependence on distance).

The total free parameters are $\{M_{\text{tot}},c_{0.1},\Upsilon_{\text{disc}},\Upsilon_{\text{bulge}},{i},{D}\}$. The priors on $\Upsilon_{\text{disc}}$ and $\Upsilon_{\text{bulge}}$ are lognormal with means of $\log(0.5)$ and $\log(0.7)$ respectively, and 0.1 dex scatter \citep[following][]{liComprehensiveCatalogDark2020}. The priors on $i$ and $D$ are normal with mean given by the observed values and scatter by the observational uncertainties from SPARC. The posterior is sampled using the \texttt{emcee} ensemble sampler \citep{foreman-mackeyEmceeMCMCHammer2013}. We set the number of walkers to 200 and the stretch move to $a=2$. To ensure the chain is converged we run the sampler until the chain is at least 50 times the autocorrelation length \citep{goodmanEnsembleSamplersAffine2010} in all parameters, or a minimum of 10,000 steps to ensure the posterior is densely sampled to aid in the calculation of $\dkl$. The first 25 autocorrelation lengths are discarded as burn-in.

\subsection{Goodness-of-fit}\label{sec:residuals}

 We wish to examine the dependence of constraining power on the type and precision of the measurements. A nuisance effect is that constraints can be very tight for models that are a poor fit to the data, as a small fraction of parameter space can still have high likelihood relative to the rest of it, even if the absolute value of the likelihood is low for that region (a problem previously identified for rotation curves, e.g. \citealt{liComprehensiveCatalogDark2020}).
 
 To exclude galaxies that are poor fits to a particular profile, we examine the distribution of normalised residuals

\begin{equation}\protect{}{ 
\mathcal{R}_i = \frac{
V_{\text{i,obs}} - V_{\text{pred}}(r_\text{i}) }{ \delta V_{\text{i,obs}} }},
\end{equation}
evaluated for the $i$ RC datapoints of a galaxy. The set of $\mathcal{R}_i$ should be drawn from a standardised normal distribution if the model is perfect \citep{andraeDosDonTs2010,zentnerCriticalAssessmentSolutions2022}. We identify galaxies as having poor fits if the probability that the distribution of residuals is drawn from a standardised normal is $p_\text{fit} < 0.05$, as calculated by the Kolmogorov–Smirnov test \citep{masseyKolmogorovSmirnovTestGoodness1951}. The probabilistic nature of the test means it is more stringent for better sampled rotation curves, which is desirable as better sampled rotation curves generally give stronger constraints on halo properties. We define a galaxy as overfit if $p_\text{fit} < 0.05$ and the standard deviation of their residuals is less than 1, and underfit if $p_\text{fit} < 0.05$ and the standard deviation of residuals is greater than 1. This procedure finds 23 (14) galaxies to be underfit and 8 (11) overfit for the NFW (Burkert) profile in the fiducial model. These are removed from the sample.

We find that the galaxies for which Burkert is underfit have higher than average mass, but for NFW there is no clear trend with any galaxy property (including inclination). There is no clear trend for overfitting using either halo profile. Finally, removing under/overfit galaxies does not significantly impact the distribution of $\dkl$ for the sample. This suggests that whether or not the above procedure has identified all poor fits, the issue of poorly fitting galaxies having tight constraints is unlikely to bias our subsequent analysis of $\dkl$.

\subsection{Abundance matching}

For reference we also show the information gain on halo properties from abundance matching, an empirical model that matches the haloes in simulations to observed galaxies by positing an approximately monotonic relationship between a halo property (the \emph{proxy}) and a galaxy observable (e.g. \citealt{kravtsovDarkSideHalo2004,conroyModelingLuminosityDependentGalaxy2006}). In the simplest model the proxy is halo mass and the galaxy observable is stellar mass or luminosity. We use the proxy of \cite{lehmannConcentrationDependenceGalaxyHalo2016} which models assembly bias through the hybrid proxy $v_\beta:=V_{\mathrm{halo}}\left(\frac{v_{\max}}{V_{\mathrm{halo}}}\right)^\beta$, where $v_{\text{max}}$ is the maximum circular velocity of the halo, and $V_{\text{halo}}$ is the velocity at the virial radius. The free parameters in the model are the AM scatter $\sigma_{\text{AM}}$ and $\beta$, which they constrain by clustering to $\beta=0.57^{+0.20}_{-0.27}$ and $\sigma_{\text{AM}}=0.17^{+0.03}_{-0.05}$.  We calculate the posterior on mass--concentration using the Bayesian inverse subhalo abundance matching scheme of \cite{yasinInferringDarkMatter2022} (section 3.3), sampling over $\sigma_{\text{AM}}$ and $\beta$.

\subsection{The Kullback--Leibler divergence}\label{sec:kl}

The Kullback--Leibler divergence \citep{kullbackInformationSufficiency1951} can be used to quantify the information gain in an experiment in going from the prior distribution to the posterior in units of bits

\begin{equation}
D_\text{KL}(P \parallel \pi) = \int_\theta \mathcal{P}(\theta) \log_2\left(\frac{\mathcal{P}(\theta)}{\pi(\theta)}\right)\ \operatorname{d}\!\theta .
\end{equation}
It quantifies the similarity between $\mathcal{P}(\theta)$ and the reference distribution $\pi(\theta)$. In information theory terms it is the excess surprise when using $\mathcal{P}(\theta)$ compared to $\pi(\theta)$. It is the appropriate metric to use when comparing the improvement on precision in constraints between two experiments \citep{buchnerIntuitionPhysicistsInformation2022a}, as it takes into account the full probability distributions, as opposed to comparing a summary statistic such as the $2 \sigma$ credible interval. However this comes at the expense of ease of interpretation. An intuitive example is an experiment with a flat prior that produces a flat posterior that has a $k$ times smaller hypervolume. In this case $\kl = \log_2(k)$.

\begin{figure}
\includegraphics[width=8 cm]{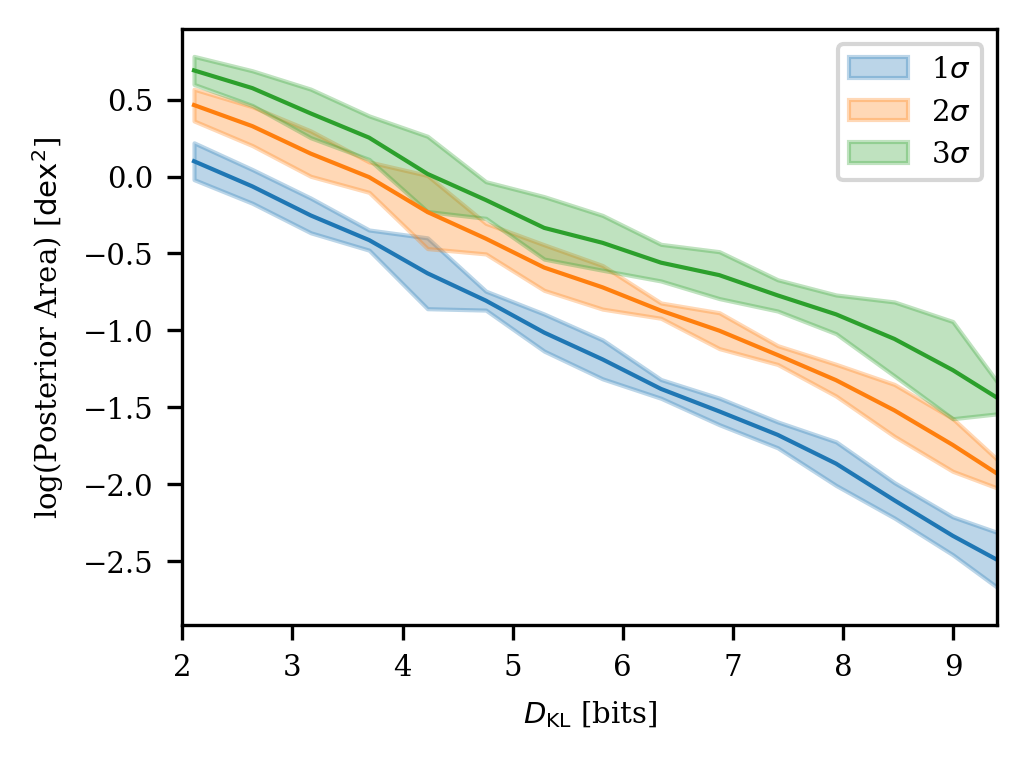}
\caption{The correlation between the KL divergence ($\dkl$) of the posterior from the prior in $M_{\textrm{tot}}-c_{0.1}$ space and the area of the 2D posterior (for varying confidence intervals) from fitting Burkert haloes to the full RC for the whole sample of galaxies. The solid lines show the mean posterior area in bin of $\dkl$ (15 bins total), with the band showing the $1\sigma$ spread in each bin. $\dkl$ and posterior size are strongly correlated, with a small scatter due to changing shapes of the posteriors and the prior on $M_{\textrm{tot}}$ that is a function of the galaxy's $\mbar$.}
\label{fig:kl}
\end{figure}

To calculate $\dkl$, kernel density estimation (KDE) is used to estimate the posterior probability distribution. We use the fastKDE algorithm \citep{obrienReducingComputationalCost2014,obrienFastObjectiveMultidimensional2016}, which selects the kernel and bandwidth based on the criteria of \citet{bernacchiaSelfconsistentMethodDensity2011}. We ensure that $\dkl$ is converged with respect to the number of MCMC samples by checking that our results do not change when using a shorter or longer chain.

In Fig.~\ref{fig:kl} we plot $\dkl$ against the 2D credible interval size for one of our runs, showing the strong correlation between them. Two factors cause a scatter between $\dkl$ and the size of a contour. Firstly two posteriors with the same size 2$\sigma$ contours will have a different $\dkl$ if the rest of their contours are different. The prior on $M_{\text{tot}}$ is also a function of the galaxy's $\mbar$.  $\dkl$ is dependent on the size of the prior. Our prior bounds are well motivated for mass, and the lower bound of concentration. However the upper bound of concentration is arbitrary. However we are interested in the relative $\dkl$ between different measures, so our conclusions are not sensitive to the choice of prior.

\subsection{A predictive model for $\dkl$}    
\label{sec:r2}

We aim to study the dependence of $\dkl$ on the properties of the measurement and the properties of the galaxy. To do this we build a predictive model for $\dkl$  using the \texttt{ExtraTrees} algorithm \citep{pedregosaScikitlearnMachineLearning2011}. We optimise the hyperparameters through a grid search with 5-fold cross-validation \citep[see][]{kohaviStudyCrossValidationBootstrap1995}. The features are the set of galaxy properties listed in the SPARC database. In addition, when fitting using the full RC, we add the following features that describe the details of the galaxy and RC: $N$ ( the number of RC data points); $\rout$ (the radius of the outermost data point); $\rout/R_{\text{eff}}$ (the ratio of the radius of the outermost data point to the effective radius of the galaxy); $\frac{1}{N} \sum_i V_{\text{obs,i}} / V_{\text{bar,i}}$ (the mean ratio of the observed velocity to the baryonic circular velocity, which quantifies the mean dark matter dominance); $V_{\text{obs}}(r) / V_{\text{bar}}(r)$ at the radii $R_{\text{out}}$, $R_{\text{disk}}$ and $R_{\text{2.2}}$ (this quantifies the dark matter dominance at different points in the galaxy); the summary statistics and their uncertainties; $\delta V_{\text{out}} / V_{\text{out}}$ (the uncertainty on the outermost data point); $\frac{1}{N} \sum_i \delta V_{\text{obs,i}} / V_{\text{obs,i}}$ (the mean velocity uncertainty). When fitting using individual summary statistics, as most of the above features are not relevant, we only add the summary statistic and its uncertainty, as well the ratio of the summary statistic to the baryonic circular velocity at the corresponding radius e.g. $\vmax / V_{\text{bar}}(R_{\text{max}})$.

To find which features are most important in determining $\dkl$, we use the feature importance analysis method of \citet[section~3.6]{stiskalekScatterGalaxyhaloConnection2022}. Features are added to the list of features used to train the \texttt{ExtraTrees} regressor one at a time, with the feature added at each increment the one that yields the greatest improvement in accuracy. This produces a list of features, ranked from most important (added first) to least important (added last), and the new accuracy after their inclusion. This method avoids the ambiguities associated with correlated features. Due to the small sample size, we divide the sample into 10 and calculate predictions for each subsample using a regressor trained on the rest of the samples. The accuracy of the model's predictions are assessed using the coefficient of determination \citep{draperAppliedRegressionAnalysis1998}

\begin{equation}
R^{2}=1-\frac{\sum_{i}(y_{i,\text{true}}-y_{i,\text{pred}})^{2}}{\sum_{i}(y_{i,\text{true}}-\hat{y}_{\text{true}})^{2}},
\label{eq:r2}
\end{equation}
where $y_{i,\text{true}}$ is the test set value, $y_{i,\text{pred}}$ the corresponding prediction and $\hat{y}_{\text{true}}$ the mean test set value. $R=1$ corresponds to perfect accuracy, and $R=0$ to a model that always predicts $\hat{y}_{\text{true}}$ irrespective of the data.

\begin{figure}
    \includegraphics[width=8 cm]{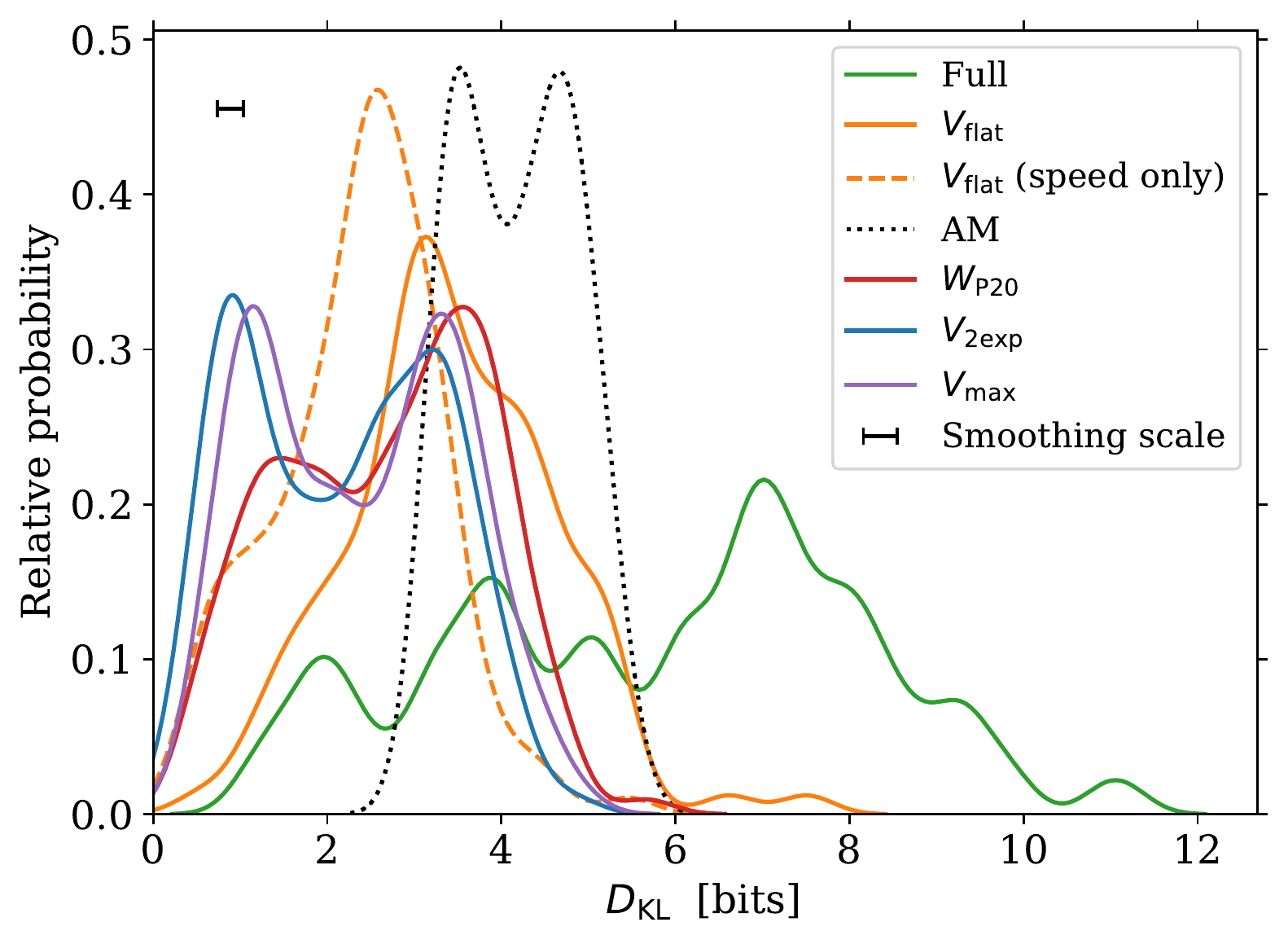}
    \caption{Kernel density estimation plots showing the distribution of the Kullback--Leibler divergence ($\dkl$) for the different types of measurement, smoothed with a Gaussian kernel with a standard deviation of 0.3 $\dkl$ to remove unphysical small-scale noise. The full rotation curve (``Full'') contains the most information. The single point summary statistics  ($\wpt$/$\vmax$/$\vexp$) contain much less information and are similar to each other. $\vflat$ has much less information than the full RC, albeit more than other summary statistics, showing the importance of the inner parts of the RC in constraining the shape of the halo and breaking the degeneracy between mass and concentration. Abundance matching has more information than any measure except the full RC.}
    \label{fig:statistics}
\end{figure}

\begin{figure*}
\centering
\includegraphics[width=\textwidth]{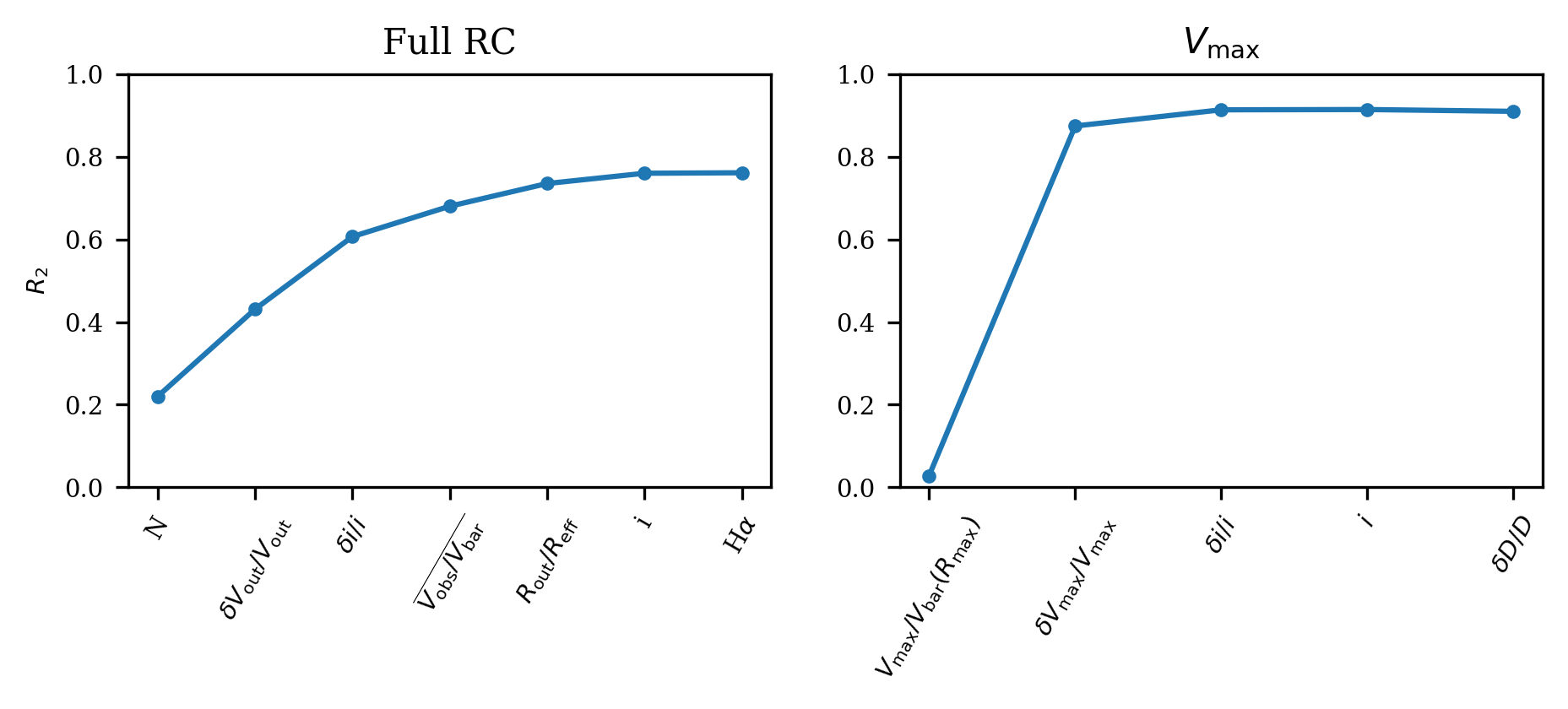}
\caption{The important features for predicting the KL divergence ($\dkl$) from fitting to the full rotation curve (\emph{left panel}) and fitting to $\vmax$ (\emph{right panel}) using an ExtraTrees regressor. Features are added left to right in the order which maximises accuracy at each increment, as described in Section \ref{sec:r2}. No features are predictive on their own, but for the full rotation curve the combination of number of data points $N$, the fractional uncertainty on the outermost measured velocity $\delta \vout / \vout$, the fractional uncertainty on inclination $\delta i / i$, and the mean ratio of observed rotational velocity to baryonic circular velocity $\overline{V_{\text{obs}}/ V_{\text{bar}}}$ (a measure of dark matter dominance) give reasonable accuracy, with $R_2=0.77$. For $\vmax$ a combination of $\vmax / V_{\text{bar}}(\rmax)$  and the fractional uncertainty $\delta \vmax / \vmax$ give good accuracy ($R^2=0.9$). The full list of features used in our analysis are described in Sec.~\ref{sec:r2}, but includes all galaxy properties (such as stellar mass) given in the SPARC database, as well as additional features characterising the rotation curve (such as H$\alpha$, a binary variable for whether or not a galaxy has H$\alpha$ kinematic data.)}
\label{fig:full_r2}
\end{figure*}

\begin{figure*}
\centering
\includegraphics[width=\textwidth]{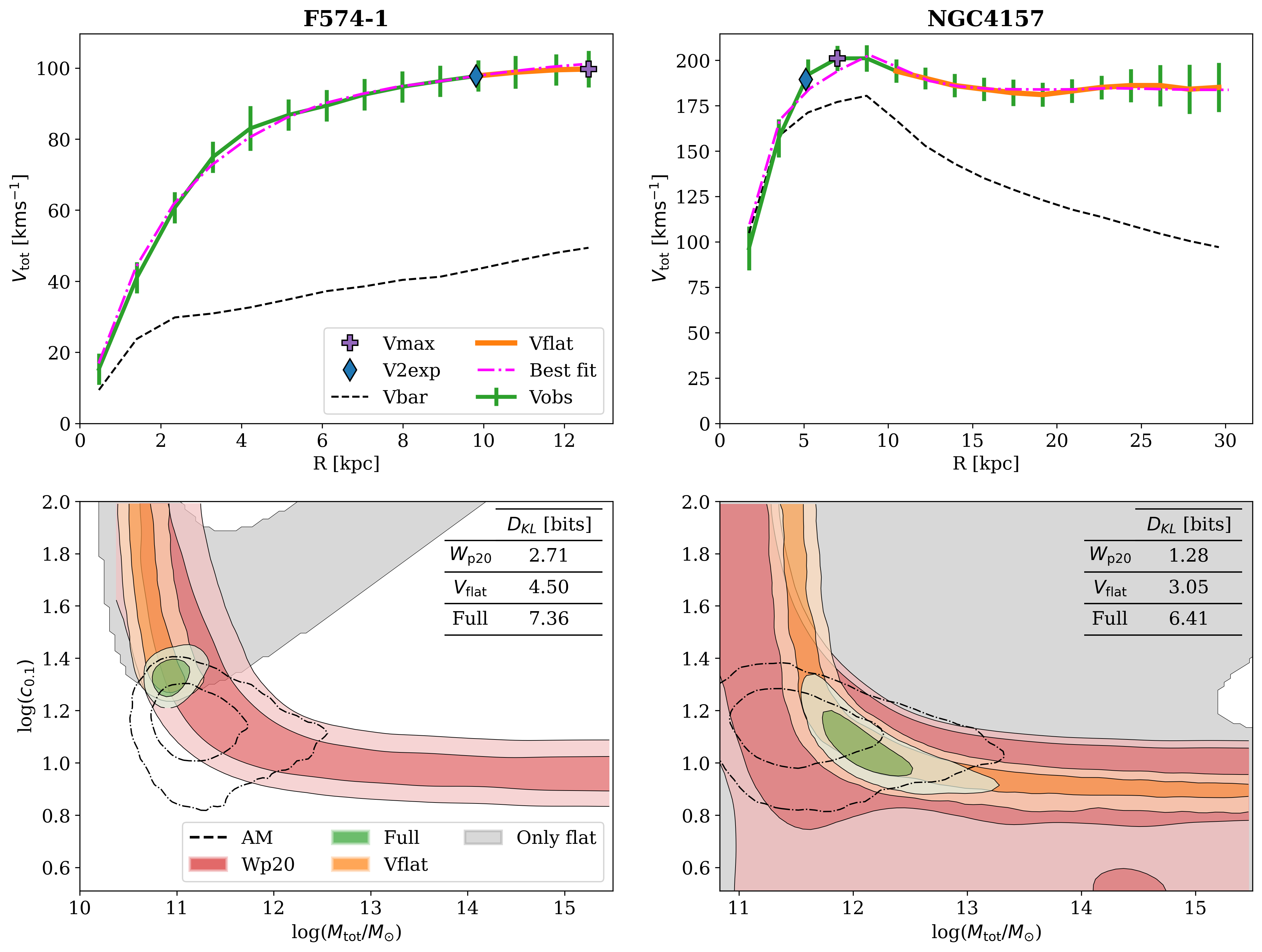}
\caption{Comparison between observed and model rotation curves (\emph{top panels}), and the posteriors in mass--concentration space for F574-1 (\emph{left panel}), a low-surface brightness galaxy, and NGC4157  (\emph{right panel}), an intermediate mass spiral galaxy. Different colours in the bottom panel show posteriors computed from fitting to the full rotation (green), fitting to the HI linewidth $\wpt$ (red) and fitting to the $\vflat$ summary statistic (orange, see Section \ref{sec:vflat}). The grey shows the posterior from just requiring that the rotation curve flatness condition be met, without matching the actual value of $\vflat$. The $\dkl$ of each posterior is shown on the right. The posteriors for $\vmax$ and $\vexp$ are not shown, but are similar to $\wpt$. The abundance matching posterior for each galaxy is also shown for comparison.}
\label{fig:posteriors}
\end{figure*}

\section{Results}\label{results}

\subsection{Summary statistics}\label{sec:summ_stat}

\subsubsection{Overview} 
\label{sec:full}

In our primary analysis we wish to study the dependence of the KL divergence ($\dkl$) on the type of measurement, properties of the measurement, and properties of the galaxy. In Fig.~\ref{fig:statistics} we show the distribution of $\dkl$ when fitting to the different types of measurement in the kinematic inference: the full RC, $\wpt$, or the summary statistics (see Table \ref{tab:definitions}). The full rotation curve produces the tightest constraints, with a fairly flat distribution of $\dkl$ between 4 and 10 bits (corresponding roughly to posteriors that are 16 and 1000 times smaller than the prior). The broad distribution of $\dkl$ is due to the massive variation both in measurement properties and galaxy properties across the sample. The summary statistics and $\wpt$ produce similar distributions in $\dkl$, with  a minimum of close to 0 bits (meaning the posterior is similar to the prior) and a maximum of ~6 bits (approximately half that of the full RC fits). $\vflat$ has a higher mean $\dkl$ than the other summary statistics. We also show the distribution of $\dkl$ from abundance matching, which is narrow and has a mean 1-2 bits higher than the summary statistics.

We use the $\texttt{ExtraTrees}$ algorithm to construct a predictive model for $\dkl$ and carry out a feature importance analysis (as described in Section~\ref{sec:r2}) and show the results in Fig.~\ref{fig:full_r2}. The model is moderately predictive, with an accuracy of $R_2=0.77$. The most important features for predicting $\dkl$ are, in descending order of importance: the number of data points $N$, the uncertainty on the outermost measured velocity $\delta \vout / \vout$, the fractional uncertainty on inclination $\delta i / i$, the mean ratio of the observed velocity to baryonic circular velocity $\overline{V_{\text{obs}}/V_{\text{bar}}}$ and the ratio of the radius of the outermost data point to the effective radius $\rout/\reff$. We reiterate that the uncertainties on $V_{\text{obs}}$ do not include a contribution from inclination - which is treated separately.

As illustrative examples, in Fig.~\ref{fig:posteriors} we show the RCs and posteriors of  F574-1, a low surface-brightness galaxy, and NGC4157, an intermediate mass spiral galaxy. F574-1 is an example of a galaxy that is dark matter dominated. Its RC gradually rises and levels off to a flat part close to the last measured point. NGC4157 is baryon-dominated in its inner parts, with a RC that sharply rises to a maximum velocity that corresponds to the peak in the stellar velocity, before declining slightly to the flat part. Both galaxies have a similar number of data points, and uncertainties on observed distance and inclination. In general dark matter dominated galaxies have tighter constraints on halo properties, as when $\vbar$ is low relative to $\vobs$, the uncertainties on $\vbar$ (which are set by the uncertainties on the mass-to-light ratios) are less important. The RC of F574-1 is also sampled further out into the halo (relative to the virial radius) than NGC4157, contributing to its tighter constraints.

We note that stellar mass and surface brightness, which were input features to our feature importance analysis, do not appear among the features identified as important for predicting $\dkl$. This is because, although they are correlated with the dark matter fraction, they do not themselves directly impact the strength of the constraints on halo properties. Once $\overline{V_{\text{obs}}/ V_{\text{bar}}}$ (when fitting to the full RC) or $\vmax / V_{\text{bar}}(\rmax)$ (when fitting to $\vmax$) are selected, adding stellar mass or surface brightness does not improve predictivity further.

\subsubsection{$\wpt$, $\vexp$, $\vmax$}

Using either $\wpt$, $\vexp$ and $\vmax$ in the inference produces posteriors that are very similar in shape for most galaxies. We show the posteriors for $\wpt$ in Fig.~\ref{fig:posteriors}. As we have assumed the halo is spherically symmetric, the circular velocity due to the halo depends solely on its enclosed mass. This results in a complete degeneracy between the halo mass and concentration for $\wpt$/$\vexp$/$\vmax$, which do not constrain the shape of the RC. For the dark matter dominated F574-1, the posteriors are simply a band corresponding to the additional circular velocity required from the dark matter to generated the observed $\wpt$/$\vexp$/$\vmax$, thickened by the its observational uncertainty and the uncertainties on $i$, $D$ and $\Upsilon_{\text{disk/bulge}}$. 

For NGC4157 the constraints on halo properties from $\wpt$/$\vexp$/$\vmax$ are extremely weak, as the baryons alone can generate the observed values of these summary statistics. Therefore a large range of haloes are compatible with observation, as long as they do not significantly change: the mass enclosed within the \HI{} disk for $\wpt$; the maximum observed velocity for $\vmax$; the velocity at $R_{2.2}$ for $\vexp$. These three criteria result in similar constraints: a halo must have mass or concentration low enough such that there is no significant halo mass at lower radii. In the case of $\wpt$, the degeneracy between mass and concentration can be broken by fitting the full \HI{} flux profile rather than just the linewidth, as an extended, flat RC produces a very different \HI{} profile to a RC that peaks and then declines (as occurs with very low mass / concentration haloes). We leave this to future work.

The mean $\dkl$ for $\wpt$/$\vmax$/$\vexp$ is 2.76/2.41/2.27. In Fig.~\ref{fig:full_r2} we present the $\dkl$ feature importance analysis for $\vmax$ only. $\vmax$ is chosen because it is available for more galaxies than $\wpt$ and $\vexp$, but all three give similar results. The important features are, in descending order:  $\vmax / V_{\text{bar}}(R_{\text{max}})$ (which measures the dark matter-dominance at $\rmax$), $\delta V_{\text{max}} / \vmax$, its fractional uncertainty, and $\delta i / i$. We interpret the ordering of the mean $\dkl$ for the three measurements as being due to the dark matter dominance of the region probed by each quantity. The $\HI{}$ disc extends beyond $\rexp$, and so probes the RC in the more dark matter-dominated outer regions. For most galaxies in our sample, $\vmax$ coincides with the outer point of the RC (as in F574-1), which is typically beyond $\rexp$. However for the galaxies with baryon-dominated inner regions such as NGC4157 $\vmax$ roughly coincides with $\rexp$. Hence the mean $\dkl$ for $\vmax$ is between $\wpt$ and $\vexp$.

\subsubsection{$\vflat$}

$\vflat$ has higher mean $\dkl$ than $\wpt$, $\vmax$ and $\vexp$. Its posteriors (see Fig.~\ref{fig:posteriors}) are either a band similar to $\wpt$ (NGC4157) or a truncated band (F574-1). There are two distinct contributors to the constraining power on halo properties from the $\vflat$ statistic. The first is that $\mvir$ and $c_{0.1}$ must generate a RC that meets the flatness criterion. If this is met, then the velocity of the flat part of the model RC must also be equal to the observed $\vflat$. The second criteria is similar to the summary statistics $\vexp$ and $\vmax$, in that it simply requires one part of the RC to be a certain value, resulting in a degenerate band posterior. It is the flatness criterion that can truncate the band, as it does for F574-1.

To demonstrate the behaviour of the flatness criterion in isolation, in Fig.~\ref{fig:posteriors} we show the regions for which the RC is considered flat in grey. The shape of the RC depends on the parameters describing the baryons, so we consider a given $M_{\text{tot}},c_{0.1}$ to meet the flatness criterion if the probability of the RC being flat is $>34\%$ (i.e. 1$\sigma$)
when marginalising over $D$ and $\Upsilon$. A much smaller region of the $M_{\text{tot}},c_{0.1}$ prior is considered flat for F574-1 than NGC4157. This is because for NGC4157 the RC with baryons alone is considered flat, as it is gently declining over many disc scale lengths. But generating a flat RC for F574-1 requires a dark matter halo that is both dominant over the baryonic component and has high enough concentration to have reached the gently declining "flat" part by the outer most RC point.

To separate the second criteria out from the flatness requirement, we calculate $\dkl$ for a new summary statistic: \emph{$\vflat$ (speed only)}. This is a single data point that is the mean speed of the flat part of the RC, occurring at its mean radius, without any flatness requirement. We see in Fig.~\ref{fig:statistics} that the $\dkl$ for $\vflat$ (speed only) are similar to $\wpt$,$\vmax$ and $\vexp$. This exercise demonstrates the extra constraining power that comes from observing a flat rotation curve over its length, compared to just measuring a single point from it.

We train an \texttt{ExtraTrees} regressor on $D_{\text{KL}}$ for $\vflat$, but found it to be poorly predictive. This is due to difficulty in predicting the size of the region for which the flatness criterion is met, which depends on the detailed shape of the circular velocity due to the baryons.

\subsection{$\dkl$ as a function of measurement properties}

\begin{figure}
\hypertarget{fig:sd_}{%
\centering
\includegraphics[width=8 cm]{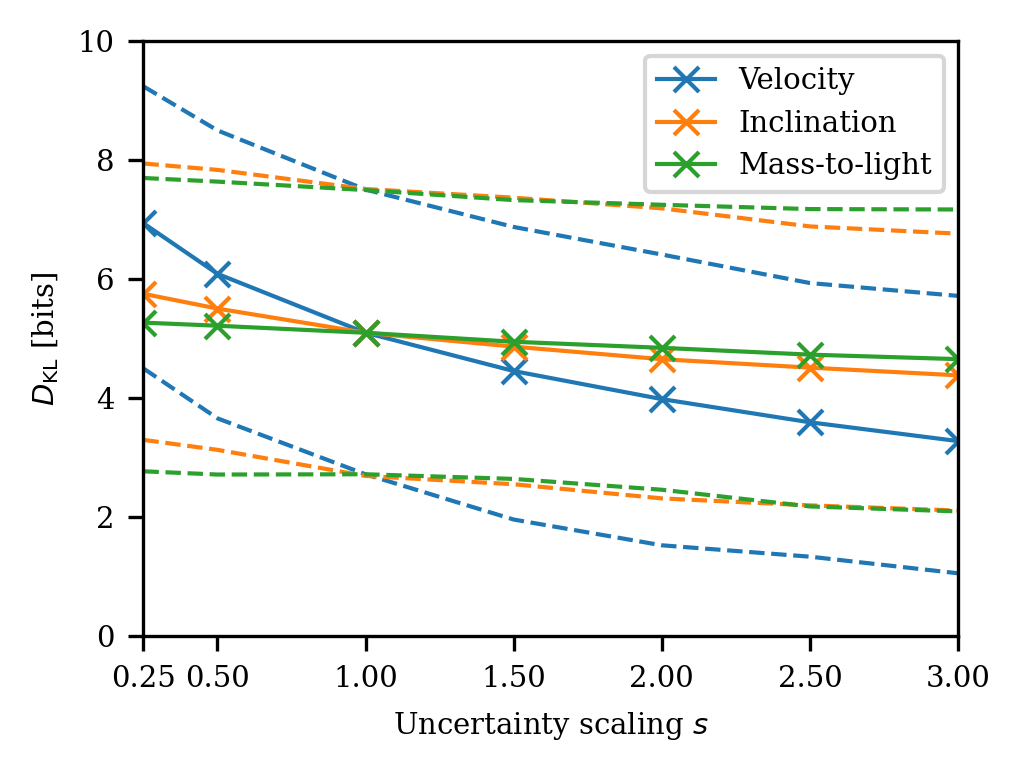}
\caption{The dependence of $\dkl$ on the uncertainties on velocity, inclination and mass-to-light ratio. At each point, the uncertainties are scaled on either the velocities, the inclination, or the mass-to-light ratios on the bulge and disk by a constant factor $s$ (such that $\delta ' = s\delta $) and the $\dkl$ from fitting to the full rotation curve is recalculated for each galaxy. Any galaxy that is considered underfit (see \ref{sec:residuals}) for any value of $s$ for any of velocity/inclination/mass-to-light is excluded from this analysis. Therefore $\dkl$ is calculated for the same sample of 98 galaxies for all points. The solid lines show the mean of $\dkl$ at each value of $s$ (marked by crosses), and the dashed lines show the 16th and 84th quantiles of the distribution. $\dkl$ is most dependent on the velocity uncertainties.}}
\label{fig:scale}
\end{figure}

In the feature importance analysis the uncertainties on inclination and velocity were found to be important predictors of $\dkl$. We now isolate their effect on $\dkl$ for the full RC by scaling their uncertainties by a constant factor $s$, i.e.
$\delta V_{\text{obs,scaled}} = s \delta V_{\text{obs}}$ or $\delta i_{\text{obs,scaled}} = s \delta i_{\text{obs}}$,
and repeating the inference. We do this for $\delta\vobs$ and $\delta i$ separately, for a range of values of $s$. We also apply the same procedure to the scatter on the prior of $\Upsilon_{\mathrm{disc/bulge}}$ (changing it for both disc and bulge simultaneously).

We exclude bad fits using the residual analysis described in Section \ref{sec:residuals}. To ensure we use the same sample for all three quantities, galaxies are only included if they are not bad fits for any value of $s$ for all of $\vobs$, $i$ and $\Upsilon_{\text{disc/bulge}}$. With a minimum value of $s=0.25$ (the fits are worst for lower $s$), this leaves 98 galaxies in the sample. We show their $\dkl$ as a function of $s$ in Fig.~\ref{fig:scale}. $\dkl$ shows the greatest dependence on the velocity uncertainties, and is relatively flat for the rest. The rate of increase in $\dkl$ steepens as $s$ decreases for $\delta V_{\text{obs}}$ and $i$.

\begin{figure*}
\centering
\includegraphics[width=\textwidth]{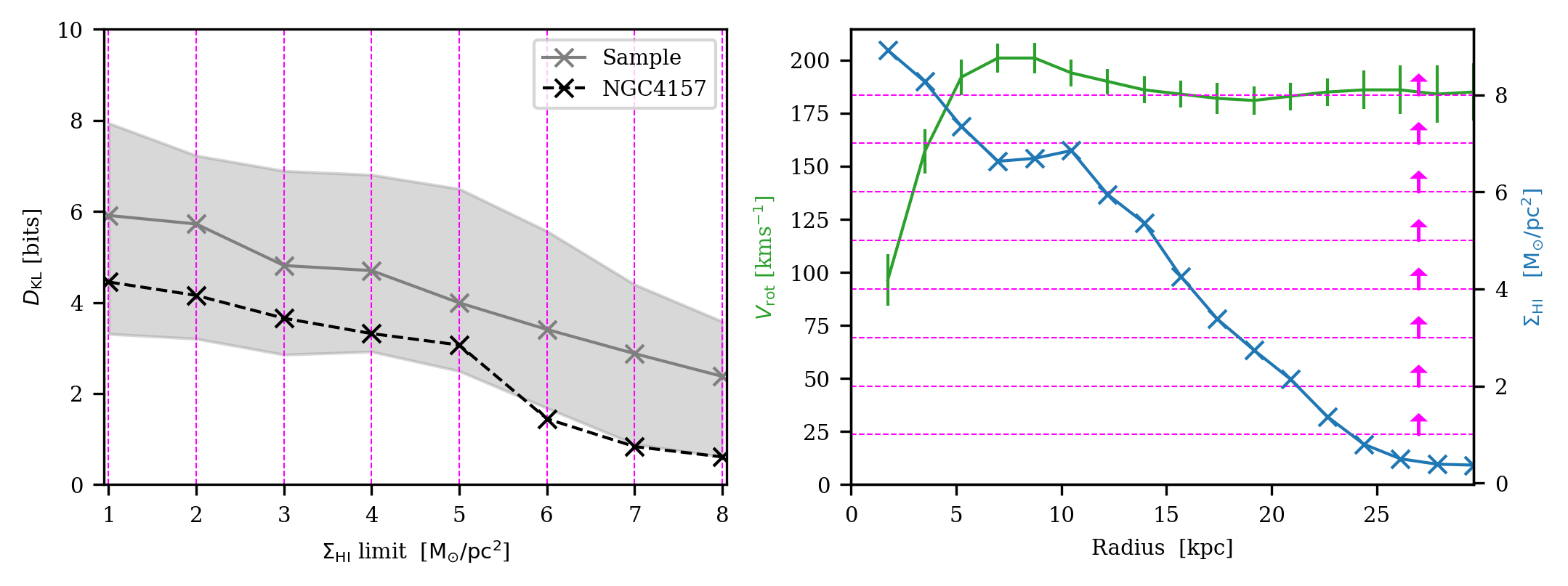}
\caption{The dependence of $\dkl$ (grey line, sample mean; band, 16th and 84th quantiles) on the minimum HI surface density probed $\Sigma_{\text{HI,min}}$ (left panel). The procedure for varying $\Sigma_{\text{HI,min}}$ is illustrated for a single galaxy (NGC4157) in the right panel: for each value of $\Sigma_{\text{HI,min}}$ (pink lines), only the data points of the rotation curve (blue) at radii where $\Sigma_{\text{HI}}$ > $\Sigma_{\text{HI,min}}$ (i.e. green > pink) are used in the inference. The $\dkl$ for NGC4137 is shown as a dashed line in the left panel. This analysis is only applied to the subsample of 54 galaxies that each have an \HI{} surface density profile that spans all the way from 1 to 8 $\Sigma_{\text{HI}} / M_{\odot}\text{pc}^{-2}$. For this sample, $\dkl$ increases strongly (with an approximately linear relationship) as $\Sigma_{\text{HI,min}}$ is reduced and the lower surface density regions towards the outskirts of the galaxy are added to the observation.}
\label{fig:sd}
\end{figure*}

$\rout$ was identified as an important feature for predicting $\dkl$. It is set by the minimum \HI{} surface density probed by the observation. We study the effect of varying the minimum \HI{} density on $\dkl$, by repeating the inference with modified RCs that only include data points at radii where the \HI{} surface density is above a chosen minimum value, which we vary. In this analysis we include only galaxies that each have a $\Sigma_{\text{HI}}(r)$ that fully spans the range 1 to 8 $\msun / \text{pc}^2$, and which do not have $H\alpha$ observations, leaving 45 galaxies. The $\dkl$ for this sample is shown in Fig.~\ref{fig:sd}. If one knew the uncertainty on $\Sigma_{\text{HI}}$, an alternative approach would be to vary the minimum signal-to-noise ratio rather than the surface density.

\subsection{Halo profile comparison}

We now study the constraining power of observations on whether a halo has a cusp or a core. In contrast to $\dkl$, we found that it was not possible to generate a decently predictive model for the precision of the inner slope constraints from the galaxy/RC features. This is likely because the relationship between galaxy/RC properties and the precision of inner slope constraints is more complicated than for $\dkl$, and so a larger sample size is required to generate a predictive model. Therefore we instead focus on analysing how varying individual features affects the the precision of the inner slope constraints. 

\begin{figure*}
\centering
\includegraphics[width=\textwidth]{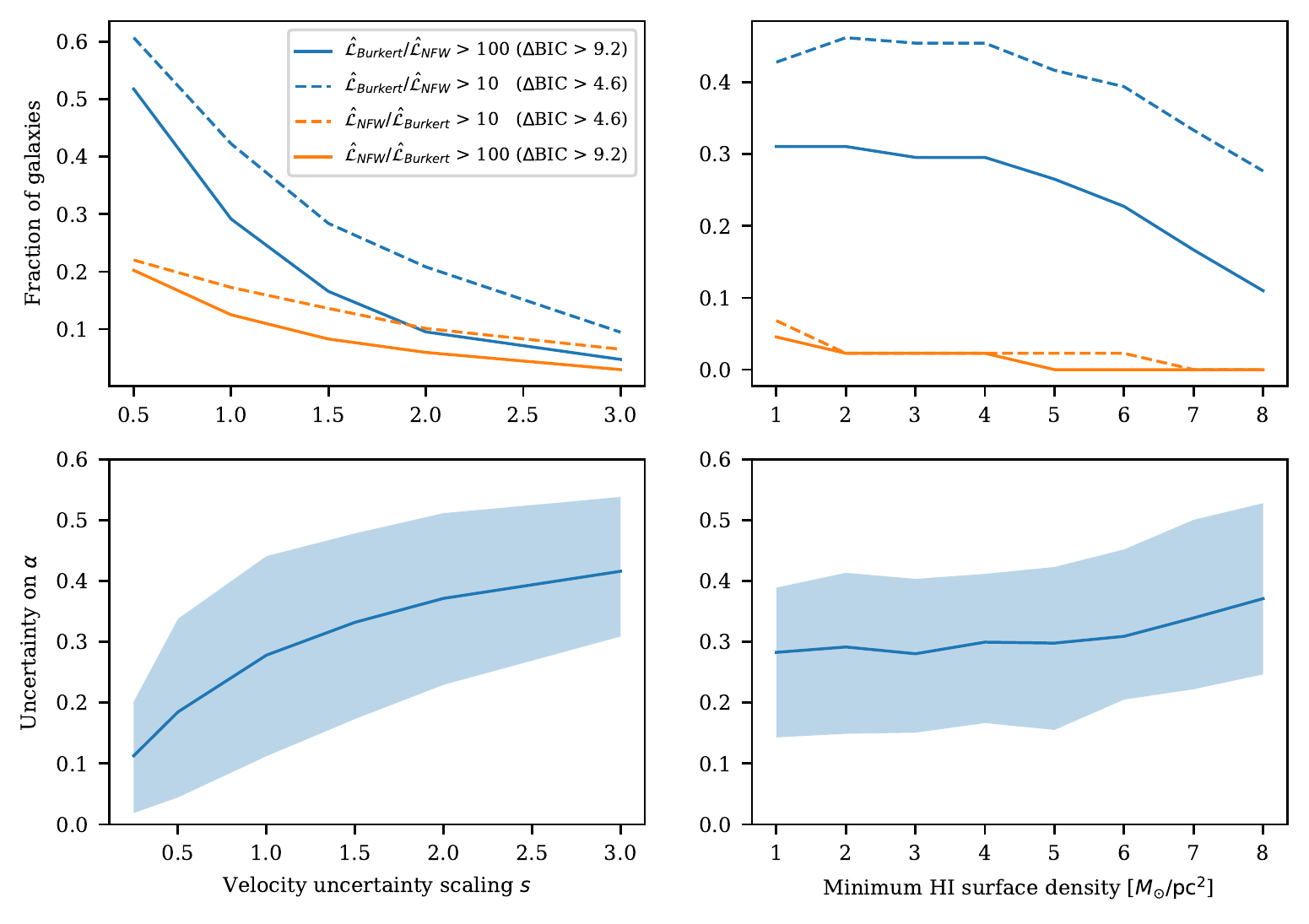}
\caption{\emph{Top panels:} The fraction of galaxies in the sample for which the ratio of the maximum likelihood estimates of the two profiles $\hat{\mathcal{L}}_{\text{Burkert}}/\hat{\mathcal{L}}_{\text{NFW}}$ (or the inverse) is above a certain value, as a function of a multiplicative scaling $s$ of the velocity uncertainties (\emph{left panels}, see Fig.~\ref{fig:scale}) and the minimum \HI{} surface density probed (\emph{right panels}, see Fig.~\ref{fig:sd}). We interpret $\hat{\mathcal{L}}_{\text{Burkert}}/\hat{\mathcal{L}}_{\text{NFW}} > 100$ (10) as strongly (moderately) favouring Burkert over NFW (and vice-versa). At $s=1$ (no scaling) one halo is strongly favoured over another in around \textasciitilde40\% of galaxies, with most favouring Burkert. If the velocity uncertainties are halved ($s=0.5$), this rises to 70\%. For reference we also show the corresponding difference in Bayesian Information Criterion.\\
%For each different line on the plot, all $\delta V_{\text{obs}}$ are scaled by a factor $s$ such that $\delta V_{\text{obs}}' = s \delta V_{\text{obs}}$, and the maximum likelihood $L$ is found for each halo profile in each galaxy. 
\emph{Bottom panels:} The distribution of the uncertainties on the $\alpha$ shape parameter from fitting a gNFW halo to the full rotation curve (solid line, mean; band, 16th and 84th quantiles in bins of $s$). Both the uncertainties on $\alpha$ and the likelihood ratios show a strong dependence on the velocity uncertainties, but a weaker dependence on the \HI{} surface density.
%  The uncertainty on the marginalised $\alpha$ parameter (which we take to be its standard deviation, std($\alpha$) measures how well the inner slope is constrained. A galaxy with a smaller uncertainty on $\alpha$ has a better known inner halo shape, with  $\Delta\alpha = 1$ the difference between a cored and a cusped ($1/r$) inner profile. We study the distribution of std($\alpha$) for the sample as a function of $s$ in Fig.~\ref{fig:discriminate
%Probing lower surface densities (below $\sim4 M_{\odot}/\text{pc}^2$) does not further constrain the inner halo shape but reducing the velocity uncertainties below the SPARC catalogue values does, showing that these should be targeted to improve our knowledge of galaxies' dark matter distributions.
}
\label{fig:discriminate}
\end{figure*}

In the $\dkl$ analysis $\delta V_{\text{obs}}$ was found to be the most important measurement uncertainty. Therefore we study the dependence of the likelihood ratio test between a NFW and Burkert profile on $\delta V_{\text{obs}}$, using the same uncertainty scaling procedure as above. We plot the resulting distribution of likelihood ratios in Fig.~\ref{fig:discriminate}. We interpret a likelihood ratio greater than 100 as one halo profile being significantly favoured over the other. For $s=1$ (no scaling) this occurs for around 30\% of galaxies, with most favouring Burkert. For $s=2$ this drops to 15\%, and for $=0.5$ it rises to ~70\%. 

Another way of looking at this is to examine how much constraining power an observation has on the shape parameter of 3 parameter halo profiles. The $\alpha$ parameter of the gNFW profile controls to the gradient of the inner slope. A cusped profile has $\alpha=1$ and a cored profile $\alpha=0$. We repeat the uncertainty scaling procedure above, but this time fitting a gNFW profile instead. The uncertainty on the marginalised $\alpha$ parameter (which we take to be its standard deviation, std($\alpha$) measures how well the inner slope is constrained. A galaxy with a smaller uncertainty on $\alpha$ has a better known inner halo shape, with  $\Delta\alpha = 1$ the difference between a cored and a cusped ($1/r$) inner profile. We study the distribution of std($\alpha$) for the sample as a function of $s$ in Fig.~\ref{fig:discriminate}. The mean scatter on $\alpha$ only drops below 0.2 for $s\sim{0.6}$. In the right column of Fig.~\ref{fig:discriminate}, we repeat the same halo profile comparison analysis as above, but this time varying the minimum \HI{} density instead $s$. We find the dependence is much weaker than for $\delta V_{\text{obs}}$.

Another possibly important factor for determining the inner slope constraints is whether or not a galaxy has $H\alpha$ observations. We split the sample in two, and find galaxies with $H\alpha$ have a mean std($\alpha$) of 0.23, but those without have a mean of 0.3. We have checked the two populations of galaxies do not significantly differ with respect to other variables that drive the constraints on $\dkl$, and so conclude that adding $H\alpha$ observations moderately reduces the uncertainty on the inner slope (comparable to halving the velocity uncertainties from their fiducial values).

% \begin{figure*}
% \centering
% \includegraphics[width=\textwidth]{images/alpha.png}
% \caption{ }
% \label{fig:alpha}
% \end{figure*}
% %We examine the effect of sensitivity on $\dkl$ rather than $\rout$, as it is the more directly controllable quantity from an observational perspective.

\section{Discussion}\label{discussion}

\subsection{Predicting information gain}

Constraining halo mass and concentration from fits to rotation curve data is routine procedure in the study of late-type galaxies. However, to our knowledge, this study is the first to formally quantify the precision of the constraints and study their variation with galaxy and measurement properties. For the SPARC sample, we found massive variation in precision on $\mtot$ and $c_{0.1}$ when fitting to the full RC, ranging from ~1 to ~11 bits of information gain. This range is equivalent to the difference between a flat prior shrinking to a flat posterior by only a factor of 2 ($2^{1}$) compared to \textasciitilde{2000} ($2^{11}$). We created a predictive model for $\dkl$ using the \texttt{ExtraTrees} algorithm, and conducted a feature importance analysis to identify the galaxy and measurement properties that are the strongest predictors of information gain. The measurement properties are, in descending order of importance: the number of data points $N$,  the fractional uncertainty on the outermost measured velocity $\delta \vout / \vout$, the fractional uncertainty on inclination $\delta i/i$, and the radius of the outermost measured velocity normalised by the effective radius $\rout/\reff$. The only important galaxy property is $\overline{V_{\text{obs}}/V_{\text{bar}}}$, a measure of the dark matter dominance. $N$ ranks more highly than both the the maximum radius of the RC (although the two are positively correlated, with a Spearman coefficient of 0.58) and whether or not a galaxy has H$\alpha$ data. This shows the importance of sampling at many points across the RC to constrain the shape.

The moderate predictivity of our model ($R^2=0.77$) is part due to the small sample size, and part due to the input features not fully capturing the full details of the RC and the distribution of the baryons. For example $N$ does not account for the autocorrelation of the RC, and $\overline{\vobs/\vbar}$ is only an average. Our model is more predictive for the fits to the single data point summary statistics, $\wpt$, $\vexp$ and $\vmax$, with the latter giving $R^2$=0.9 and depending only on the uncertainty on $\vmax$ and $\vmax/\vbar(\rmax)$, a measure of dark matter dominance.

%(although part of it is systematic, calculated from the difference in rotation speeds between the two sides of the disc).

Our feature importance analysis only ranks the measurement properties by importance. In order to quantify the size of their effect on $\dkl$, we varied the uncertainties on velocity and inclination whilst holding the rest of the inference constant. We also varied uncertainties on the mass-to-light ratios $\Upsilon_{\text{disc}}$ and $\Upsilon_{\text{bulge}}$, applied in our model through the priors. We find that the constraints are most dependent on the velocity uncertainties, with the inclination and mass-to-light showing only weaker dependence. In Fig.~\ref{fig:sd} we calculated the dependence of $\dkl$ on the minimum \HI{} density probed (which sets the maximum radius of the RC). Reducing the minimum density by $1 \msun / \text{pc}^2$ yields an additional ~0.4 bits of information gain. In order to investigate whether this dependence on the minimum \HI{} density simply mirrors the strong dependence on $N$ found earlier (Fig.~\ref{fig:full_r2}), we also plotted $\dkl$ / $N$ for the different runs (not shown), and found no strong trend with the minimum \HI{} density. This suggests that the observed trend in $\dkl$ with minimum \HI{} density is driven primarily by $N$, rather than say the points in the outer RC (where the \HI{} density is lowest) yielding disproportionately more information. This concurs with our earlier finding that $R_{\mathrm{out}}/R_{\mathrm{eff}}$ is a less important feature than $N$.

Our results can inform future survey design, by highlighting which features of the measurement should prioritised for optimisation. The main way to improve constraints is of course to use longer integration times or higher instrument sensitivity, which would increase the number of data points \citep{staveley-smithHIScienceSquare2015}. 
However specific optimisations are possible. The beam size sets the maximum resolution. At fixed sensitivity/integration time, there is a trade off between minimum \HI{} density that can be probed (i.e. the maximum radius) and resolution, although this can be altered with adaptive smoothing techniques in post-processing \citep{briggsHighFidelityDeconvolution1995}. Reducing the velocity uncertainties would require improving the model used to determine the velocities and/or increasing the spectroscopic resolution. SPARC inclinations are produced as output of the fits to the 2D velocity field \citep{lelliSPARCMASSMODELS2016}, so the uncertainties would be reduced by improving the velocity map. Inclination can also be calculated using the ellipticity of the \HI{} zeroth-moment map \citep{ponomarevaMIGHTEEHBaryonicTully2021}, including forward modelling it to the datacube \citep{mancerapinaNoNeedDark2021}, or more imprecisely using optical data. \cite{kourkchiEvaluatingSpatialInclination2022a} recently used a combination of machine learning and citizen science to improve inclinations from optical data. \cite{schombertStellarMasstolightRatios2022} use stellar population models to study the variation of the uncertainty on $\Upsilon$ with the passband, and the available morphology and colour information.

\subsection{Summary statistics}

We compared the information gain when constraining halo parameters using the full RC, summary statistics or the \HI{} linewidth $\wpt$. Our feature importance analysis found the degree of dark matter dominance at the radius of the velocity measurement was the most important factor in determining $\dkl$. The ranking of mean $\dkl$ for the summary statistics from highest to lowest is $\{\vflat,\wpt,\vmax,\vexp\}$, which reflects how far out into the halo  each measurement probes, and hence the degree of dark matter domination. L19 calculated the intrinsic scatter of the BTFRs constructed using the different summary statistics as the velocity measure. For the ones studied in this paper, they found $\{\vflat,\wpt,\vmax,\vexp\}$ had an intrinsic orthogonal scatter of $\{0.026,0.035,0.040,0.070\}$ respectively, which is the inverse ordering of $\dkl$. L19 interpret the amount of scatter to be negatively correlated with the closeness of the summary statistic to the true flat value of the rotation curve (as opposed to the observed $\vflat$, which is limited by maximum radius probed in some galaxies). Hence it is not surprising that the mean information content on halo properties from each measurement is negatively correlated with the scatter of their respective BTFRs.

In this work we chose not to apply a prior based on the mass--concentration relationship from dark matter-only simulations. Although this would have helped break the degeneracy between mass and concentration (see also Sec.~\ref{sec:introduction}), it would also have imported assumptions from $N$-body simulations and galaxy formation theory (assembly bias) which we prefer to avoid. However we now qualitatively discuss the effect this prior would have on our results. Applying the mass--concentration prior increases $\dkl$ significantly for all galaxies, with the effect greatest when using relatively weak data such as summary statistics for which the mass--concentration degeneracy is especially pronounced. Applying the prior whilst assuming a cuspy profile leads to a finite constraint on halo mass even when using such summary statistics (this is studied for linewidths in more detail in \citealt{yasinInferringDarkMatter2022}). However for many galaxies, even with the mass--concentration prior applied there is still a strong remaining degeneracy (e.g. NGC4157 in the right panel of Fig.~\ref{fig:posteriors}), especially when assuming a cored profile. Nevertheless $\dkl$ still increases significantly compared to the no-prior case, even when the degeneracy is not fully broken. It is important to bear in mind however that in this case the information gain is not purely from the kinematic data.

\subsection{Constraining the inner halo shape}

We studied the ability of observations to constraint the inner halo shape in two ways. Firstly we studied the impact of the $\delta V_{\text{obs}}$ uncertainties on the ability to distinguish between the cored Burkert profile and cusped NFW profile using a likelihood ratio test. We found that with the unmodified velocity uncertainties, one profile was decisively favoured over another in 40\% of cases, with this rising to 70\% when the uncertainties are halved. On the other hand, when the uncertainties are doubled, one halo is only strongly favoured in only 20\% of galaxies. We also studied the uncertainty of the inner slope parameter $\alpha$ when fitting a 3-parameter generalised NFW profile. With the normal SPARC uncertainties the mean std($\alpha$) is \textasciitilde{0.3} (the change in $\alpha$ is 1 between a cored gNFW profile and a normal NFW profile), but there are a significant number of galaxies with std($\alpha$) > 0.5. This suggests a survey with velocity measurements more precise than SPARC is necessary to precisely constrain the inner halo shape for whole samples of galaxies. We repeated the analysis varying the minimum \HI{} density probed, and found a much weaker dependence. This demonstrates the importance of obtaining kinematic datasets with precise velocity uncertainties when targetting the cusp-core problem \citep[see][for a review]{delpopoloReviewSolutionsCuspCore2021} relative to probing lower \HI{} surface densities.

\subsection{Comparison to literature}

\cite{saburovaInsightDarkSide2016} used a sample of 14 galaxies from the THINGS survey \citep{walterTHINGSHINearby2008} to study the size of the uncertainties on halo parameters derived from rotation curve fitting, in particular identifying the halo concentration as often poorly constrained. The main differences to this paper are our quantification of the constraining power using the Kullback--Leibler divergence, our focus on the constraining power in the mass--concentration plane rather than the uncertainty on individual parameters, our study of the constraining power as a function of measurement and galaxy properties, and our Bayesian fitting procedure that propagates the uncertainties on galaxy parameters into the constraints on halo properties. In agreement with \citeauthor{saburovaInsightDarkSide2016} we find that for many galaxies the constraining power offered by rotation curves can still be relatively weak (as evidenced by the long tail to low $\dkl$ for the full RC in Fig.~\ref{fig:statistics}).

We have highlighted the important measurement properties which should be targeted by future surveys. Identifying the observational parameters (such as integration time) required to achieve the desired measurement properties is beyond the scope of this work. Recent work has simulated spectroscopic \HI{} observations of simulated late-type galaxies \citep{omanMARTINIMockSpatially2019}, which in theory allows an end-to-end determination of the effect of observational properties such as integration time on the constraints on halo properties. However, our analysis of real observations is an important complementary approach, as simulations still struggle to produce realistic samples of rotation curves \citep{roperDiversityRotationCurves2022}. 

In light of increasingly expensive observations, but comparatively cheap computational resources, there are an increasing number of studies examining optimal observational strategies. For example, two recent studies have used the Fisher-matrix formalism to quantify the information content in stellar streams \citep{bonacaInformationContentCold2018} and the cosmic web \citep{kosticOptimalMachinedrivenAcquisition2022a} in order to identify the best observational strategy.

\section{Conclusion}\label{conclusion}

We have used the Kullback--Leibler divergence ($\dkl$) of the posterior on total mass--concentration  (where total mass is equal to the halo mass plus the galaxy mass) from the prior to quantify the gain in information obtained from spectroscopic observations of the late-type galaxies of the SPARC database. We set the observable in the kinematic inference to be either the full rotation curve, summary statistics of the rotation curve ($\vmax$, $\vexp$, $\vflat$), or the linewidth of the integrated 21-cm spectrum, $\wpt$, in order to quantify the information contained in different parts of the rotation curve and different types of measurement. Further, to determine the properties of the measurements that are most important for the information gain, we study the variation on $\dkl$ as we modify properties of the rotation curve observations such as the uncertainties on velocity or the minimum \HI{} surface density probed.  Our conclusions are as follows:

\begin{enumerate}
    \item The full RC fitting offers a wide range of information gain for the SPARC galaxies, ranging from \textasciitilde{1} bit to \textasciitilde{11} bits. This is predominantly due to the massive range in the number of data points each rotation curve has, and the large variation in velocity uncertainties.
    \item  Fits to the summary statistics of the RCs offer much smaller gains, ranging from \textasciitilde{0} to \textasciitilde{6} bits, as the posteriors are degenerate in mass--concentration and run up against the prior bounds. $\vflat$ offers a modest increase due to the flatness constraint. For most SPARC galaxies $\wpt$, $\vflat$, $\vexp$ and $\vmax$ all probe regions of the rotation curve which are dark matter dominated, and hence contain similar information on the halo. 
    \item We measured $\dkl$ as a function of the minimum \HI{} surface density probed, and the uncertainties on velocity, inclination and mass-to-light ratios. Its dependence is strongest on the minimum surface density and the velocity uncertainties. These results can be used to weigh up the increase in precision on halo constraints afforded by improving each aspect of the measurement against the associated cost.
    \item The tightness of the constraints on the inner halo shape are strongly dependent on the velocity uncertainties, but have a much weaker dependence on the minimum \HI{} surface density. This suggests that whilst both sensitivity and velocity uncertainties are important for obtaining tight constraints on halo properties, surveys specifically targeting e.g. the cusp-core problem should prioritise the latter.
\end{enumerate}

Our study has identified the most important variables for improving the constraints on dark matter halo properties from spectroscopic observations of late-type galaxies. With forthcoming instruments set to greatly enhance our ability to probe the dark matter distribution around galaxies, in terms of number of galaxies, increasing redshift and measurement precision, these results should inform future survey design to maximise the return of knowledge on the galaxy--halo connection.

\section*{Acknowledgements}

We thank Anastasia Ponomareva, Richard Stiskalek, Johannes Buchner and Jamie Bamber for useful inputs and discussion.

HD is supported by a Royal Society University Research Fellowship (grant no. 211046).

This project has received funding from the European Research Council (ERC) under the European Union's Horizon 2020 research and innovation programme (grant agreement No 693024).

For the purpose of open access, the authors have applied a Creative Commons Attribution (CC BY) licence to any Author Accepted Manuscript version arising.

%%%%%%%%%%%%%%%%%%%%%%%%%%%%%%%%%%%%%%%%%%%%%%%%%%
\section*{Data Availability}

The data underlying this article will be made available on reasonable request to the corresponding author.

% The inclusion of a Data Availability Statement is a requirement for articles published in MNRAS. Data Availability Statements provide a standardised format for readers to understand the availability of data underlying the research results describ  ed in the article. The statement may refer to original data generated in the course of the study or to third-party data analysed in the article. The statement should describe and provide means of access, where possible, by linking to the data or providing the required accession numbers for the relevant databases or DOIs.



\bibliographystyle{mnras}
\bibliography{paper} % if your bibtex file is called example.bib

%\begin{thebibliography}{99}
%\end{thebibliography}

%%%%%%%%%%%%%%%%%%%%%%%%%%%%%%%%%%%%%%%%%%%%%%%%%%

%%%%%%%%%%%%%%%%% APPENDICES %%%%%%%%%%%%%%%%%%%%%

%%%%%%%%%%%%%%%%%%%%%%%%%%%%%%%%%%%%%%%%%%%%%%%%%%

% Don't change these lines
\bsp	% typesetting comment
\label{lastpage}
\end{document}